\documentclass[aps,twocolumn, floatfix, longbibliography, superscriptaddress]{revtex4-2}

\usepackage{graphicx}
\usepackage{dcolumn}
\usepackage{bm}
\usepackage{amssymb}
\usepackage{amsmath}
\usepackage{amsthm}
\usepackage{mathrsfs}
\usepackage{commath}
\usepackage{subfigure}
\usepackage{braket}
\usepackage{float}
\usepackage{color}
\usepackage{siunitx}
\usepackage[colorlinks=true, allcolors=blue]{hyperref}
\usepackage{hyphenat}
\usepackage{footnote}
\usepackage{notes2bib}
\usepackage{xcolor}
\usepackage{threeparttable}
\graphicspath{{figures/}}

\begin{document}

\title{AI-Accelerated Qubit Readout at the Single-Photon Level for Scalable Atomic Quantum Processors}

\author{Yaoting Zhou}
\affiliation{State Key Laboratory of Quantum Optics Technologies and Devices, Institute of Opto-Electronics, Shanxi University, Taiyuan 030006, China}

\author{Weisen Wang}
\affiliation{State Key Laboratory of Quantum Optics Technologies and Devices, Institute of Opto-Electronics, Shanxi University, Taiyuan 030006, China}

\author{Zhuangzhuang Tian}
\affiliation{State Key Laboratory of Quantum Optics Technologies and Devices, Institute of Opto-Electronics, Shanxi University, Taiyuan 030006, China}

\author{Bin Huang}
\affiliation{State Key Laboratory of Quantum Optics Technologies and Devices, Institute of Opto-Electronics, Shanxi University, Taiyuan 030006, China}

\author{Huancheng Chen}
\affiliation{State Key Laboratory of Quantum Optics Technologies and Devices, Institute of Opto-Electronics, Shanxi University, Taiyuan 030006, China}

\author{Donghao Li}
\affiliation{State Key Laboratory of Quantum Optics Technologies and Devices, Institute of Opto-Electronics, Shanxi University, Taiyuan 030006, China}
\affiliation{Collaborative Innovation Center of Extreme Optics, Shanxi University, Taiyuan 030006, China}

\author{Zhongxiao Xu}
\email{xuzhongxiao@sxu.edu.cn}
\affiliation{State Key Laboratory of Quantum Optics Technologies and Devices, Institute of Opto-Electronics, Shanxi University, Taiyuan 030006, China}
\affiliation{Collaborative Innovation Center of Extreme Optics, Shanxi University, Taiyuan 030006, China}

\author{Li Chen}
\email{lchen@sxu.edu.cn}
\affiliation{State Key Laboratory of Quantum Optics Technologies and Devices, Institute of Theoretical Physics, Shanxi University, Taiyuan 030006, China}
\affiliation{Collaborative Innovation Center of Extreme Optics, Shanxi University, Taiyuan 030006, China}

\author{Heng Shen}
\affiliation{State Key Laboratory of Quantum Optics Technologies and Devices, Institute of Opto-Electronics, Shanxi University, Taiyuan 030006, China}
\affiliation{Collaborative Innovation Center of Extreme Optics, Shanxi University, Taiyuan 030006, China}

\begin{abstract}
    Quantum state readout with minimal resources is crucial for scalable quantum information processing. As a leading platform, neutral atom arrays rely on atomic fluorescence imaging for qubit readout, requiring short‑exposure, low‑photon‑count schemes to mitigate heating and atom loss while enabling mid‑circuit feedback. However, a fundamental challenge arises in the single-photon regime where severe overlap in state distributions causes conventional threshold discrimination to fail. Here, we report an AI-accelerated Bayesian inference method for fluorescence readout in neutral atom arrays. Our approach leverages Bayesian inference to achieve reliable state detection at the single‑photon level under short exposure. Specifically, we introduce a weakly anchored Bayesian scheme that requires calibration of only one state, addressing asymmetric calibration challenges common across quantum platforms. Furthermore, acceleration is achieved via a permutation-invariant neural network, which yields a 100-fold speedup by compressing iterative inference into a single forward pass. The approach achieves relative readout fidelity above 99\% and 98\% for histogram overlaps of 61\% and 72\%, respectively, enabling reliable extraction of Rabi oscillations and Ramsey interference—results unattainable with conventional threshold-based methods. This framework supports scalable, real-time readout of large atom arrays and paves the way toward AI-enhanced quantum technology in computation and sensing.
\end{abstract}

\maketitle

\section*{Introduction}  \label{Sec1}
Measurement is fundamental to how we perceive and understand the world. From the early conceptual explorations of Heisenberg~\cite{Heisenberg1927,Zeilinger:1999:foundations} and the Stern-Gerlach experiment~\cite{GerlachStern1922,Castelvecchi2022} to the verification of Bell inequality-the cornerstone of quantum information technology~\cite{Bell1964,Clauser:1972:bell,Aspect:1982:bell,Zelinger:1998:bell}, the detection of microscopic particles has profoundly shaped our grasp of the quantum world. In general quantum information processing tasks, quantum states of microscopic objects like atoms and photons are calibrated through measurement, stabilized via measurement-based feedback and error correction~\cite{Briegel:2009:measurement,TerhalQEC2015,CaiRMPQEM2023}, and ultimately read out as classical information. Fast, high-fidelity measurement is thus a capability central to quantum processors, from superconducting circuits and trapped ions to cold atoms and solid-state spins.

In particular, recent years have witnessed remarkable advances in quantum simulation and computation using neutral atom arrays~\cite{bluvstein2024logical,bluvstein2025fault,Qiao2025RydbergQS,Chen2025RydbergQS,manetsch2025tweezer}, which offers a scalable, programmable platform for quantum processor and allows precise control and detection of quantum states of individual atoms, interacting with each other through Rydberg blockade~\cite{Endres:2016:assembly,Barredo:2016:assembler,Ahn:2016:assembler,Broweays2020NP,KlausRMP2010,Urban:2009:rydberg,Gaetan:2009:rydberg}. In such a typical atomic system where qubit readout relies on atomic fluorescence imaging~\cite{Sherson:2010:fluorescence,WongCampos:2016:imaging}, a short-exposure, low-photon-count scheme is critically needed to suppress photon-scattering-induced heating and atom loss. Going beyond fundamental requirements, mid-circuit readout~\cite{ma2023mid-circuit,Lis:2023:midcircuit,Graham:2023:midcircuit} in this platform necessitates fast, reliable state detection of selected atoms in a quantum circuit, allowing measurement outcomes to directly guide subsequent gate operations or error-correction branches~\cite{bluvstein2025fault,zhou2025low}. 

\begin{figure*}[t]
	\includegraphics[width=0.92\textwidth]{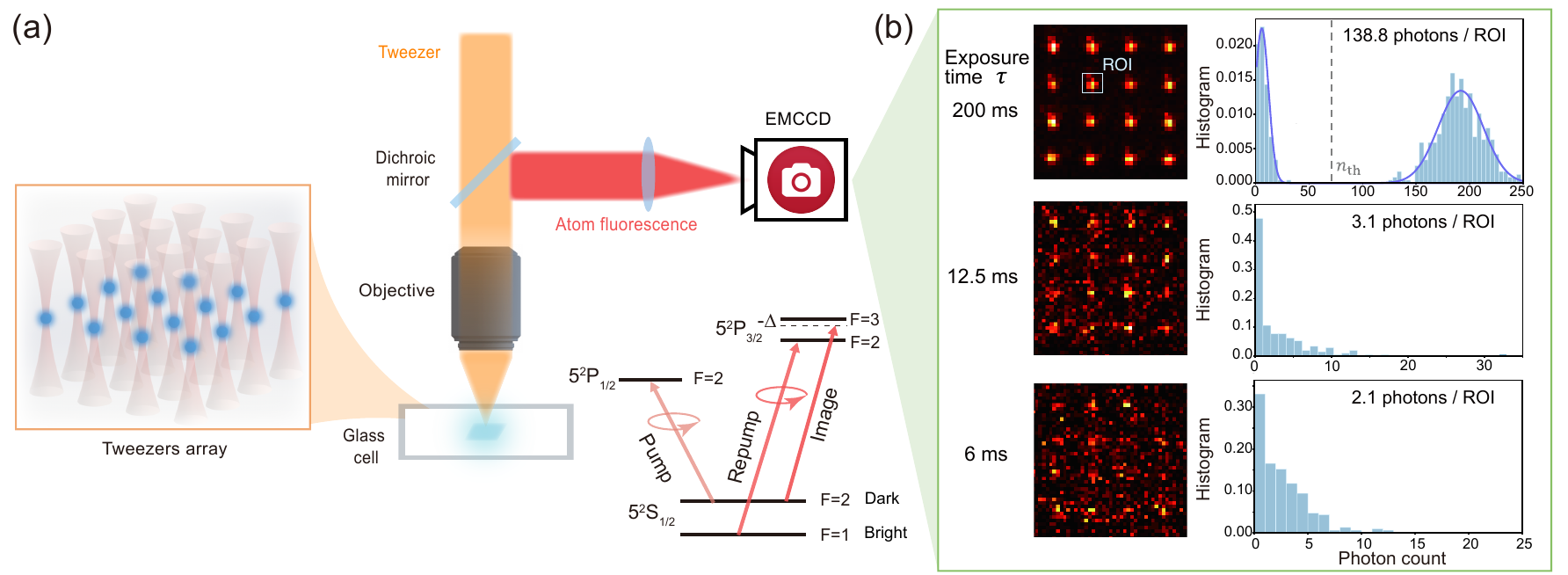}
	\caption{
	{Experimental setup and fluorescence readout.}
	(a) Schematic of the experimental apparatus showing a two-dimensional $^{87}$Rb atom array trapped in optical tweezers. Fluorescence is collected through an objective and imaged onto a EMCCD camera. Inset in the middle: energy level diagram with the dark state $|0\rangle \equiv |F=2\rangle$ and bright state $|1\rangle \equiv |F=1\rangle$. Optical pumping transfers atoms from $|1\rangle$ to $|0\rangle$ before imaging. 
	(b) Fluorescence images and photon count histograms at exposure times $\tau = $ 200 ms (top, 138.8 photons/ROI per shot), 12.5 ms (middle, 3.1 photons/ROI per shot), and 6 ms (bottom, 2.1 photons/ROI per shot). 
	At $\tau = 200$ ms, well-separated bimodal distributions (with fitting solid curves) enable conventional threshold discrimination with $n_\mathrm{th}$ (dashed line) being the threshold.
	}
	\label{fig:fig1}
\end{figure*}

However, a fundamental obstacle persists for low-photon-count readout in current array experiments. Conventional state discrimination relies on a threshold method applied to dual-peak fluorescence histograms~\cite{Endres:2016:assembly,Barredo:2016:assembler,Ahn:2016:assembler,Su:2025:imaging,Phuttitarn:2024:machinelearning}. As shown in the top panel of Fig.~\ref{fig:fig1}(b), atoms in ``bright'' and ``dark'' internal states are distinguishable, allowing an empirical threshold $n_\text{th}$ to separate the two. Yet, as the exposure time is reduced to the single-photon regime, i.e., $n\sim O(1)$, the fluorescence count distributions for the two states aggregate near zero and exhibit significant overlap (middle and bottom panels of Fig.~\ref{fig:fig1}(b))~\cite{bluvstein2024logical,manetsch2025tweezer,Todaro:2021:superconducting}. Consequently, both misclassification rates and systematic biases increase sharply, causing the threshold method to fail. The core challenge, therefore, is to extract reliable binary state information from these sparse and overlapping photon statistics.

Here, we report a scheme of fast and high-fidelity qubit readout in a scalable two-dimensional neutral atom array. Our protocol employs the Bayesian inference~\cite{VonToussaint:2011:bayesian,VanDeSchoot:2021:bayesian,Fuchs:2013:qbism,Cockayne:2019:probabilistic} to directly infer the posterior distribution from fluorescence counts, bypassing conventional threshold methods. This approach enables fluorescence readout to reach the single-photon regime per exposure. Furthermore, we accelerate the Bayesian estimation by two orders of magnitude by utilizing a \textit{permutation-invariant neural network}~\cite{Zaheer:2017:deepsets,Lee:2019:settransformer}. This network learns the mapping from photon-count data to the posterior distribution, effectively compressing the iterative Bayesian inference into a single forward propagation. We demonstrate the method on two canonical tasks: reading out Rabi oscillations and extracting decoherence times from Ramsey interferometry \cite{foot_atomic_2023}. Despite substantial histogram overlaps of 61\% and 72\%, we achieve relative readout fidelity exceeding 99\% and 98\%, respectively, inaccessible with conventional threshold methods. Notably, in contrast to strongly anchored Bayesian imaging that requires both bright and dark state distributions, we implement a weakly anchored version that uses only the dark-state distribution. This strategy is particularly advantageous for scenarios where the bright-state distribution is difficult to calibrate accurately.

\section*{Results} 
\subsection{AI-accelerated Bayesian inference} \label{Sec1}
We consider a neutral atom array with individual atoms trapped in independent optical tweezers (Fig.~\ref{fig:fig1}(a)). Quantum information is encoded in the ground-state hyperfine manifold of $^{87}$Rb atom. Concretely, the $|0\rangle$, $|1\rangle $ qubit states refer to states $|5S_{1/2}~F=2\rangle$ and $|5S_{1/2}~F=1\rangle$, and corresponds to the dark and bright states in the fluorescence readout, respectively. An arbitrary state can be expressed as $|\psi\rangle = \sqrt{1- l }|0\rangle + e^{i\phi}\sqrt{ l }|1\rangle$, with $l$ the occupation probability of the bright state and $\phi$ a relative phase. The goal is to read out the occupation $ l $ through fluorescence imaging.

Fluorescence imaging operates on the principle that, under illumination of resonant light, atoms in states $|1\rangle$ and $|0\rangle$ emit photons with different count distributions $f(n)$ and $g(n)$, with $n$ being the photon count. For a qubit with occupation $ l $, the total count 
distribution $P(n| l )$ is simply the binary mixture model, i.e., \cite{SM}
\begin{equation}
P(n| l ) = (1- l )g(n) +  l  f(n).
\label{eq:mixture}
\end{equation}
In practice, an EMCCD camera records the photon count $n$ within each region of interest (ROI) during the camera exposure time $\tau$. The distributions $f(n)$ and $g(n)$ are thus $\tau$-dependent.
As expected, at long exposure times, for instance, $\tau = 200$ ms used in our system with averaged photon count $\sim 138$ (top panel of Fig.~\ref{fig:fig1}(b)), $f(n)$ and $g(n)$ are well separated, exhibiting a pronounced bimodal structure. In this case, one can set a threshold $n_{\text{th}}$ and assign a bright-state event if $n_i > n_{\text{th}}$, and a dark-state event otherwise. The occupation is thereby estimated via the frequency of bright-state events, i.e., $ l  \approx N_>/N$, where $N_>$ and $N$ denote the number of bright-state and total events, respectively.
Such a ``threshold method" has been most extensively used for state readout in neutral atom experiments~\cite{wurtz2023aquilaqueras256qubitneutralatom,Endres:2016:assembly,Barredo:2016:assembler,Ahn:2016:assembler}. 

\begin{figure*}
	\includegraphics[width=0.8\textwidth]{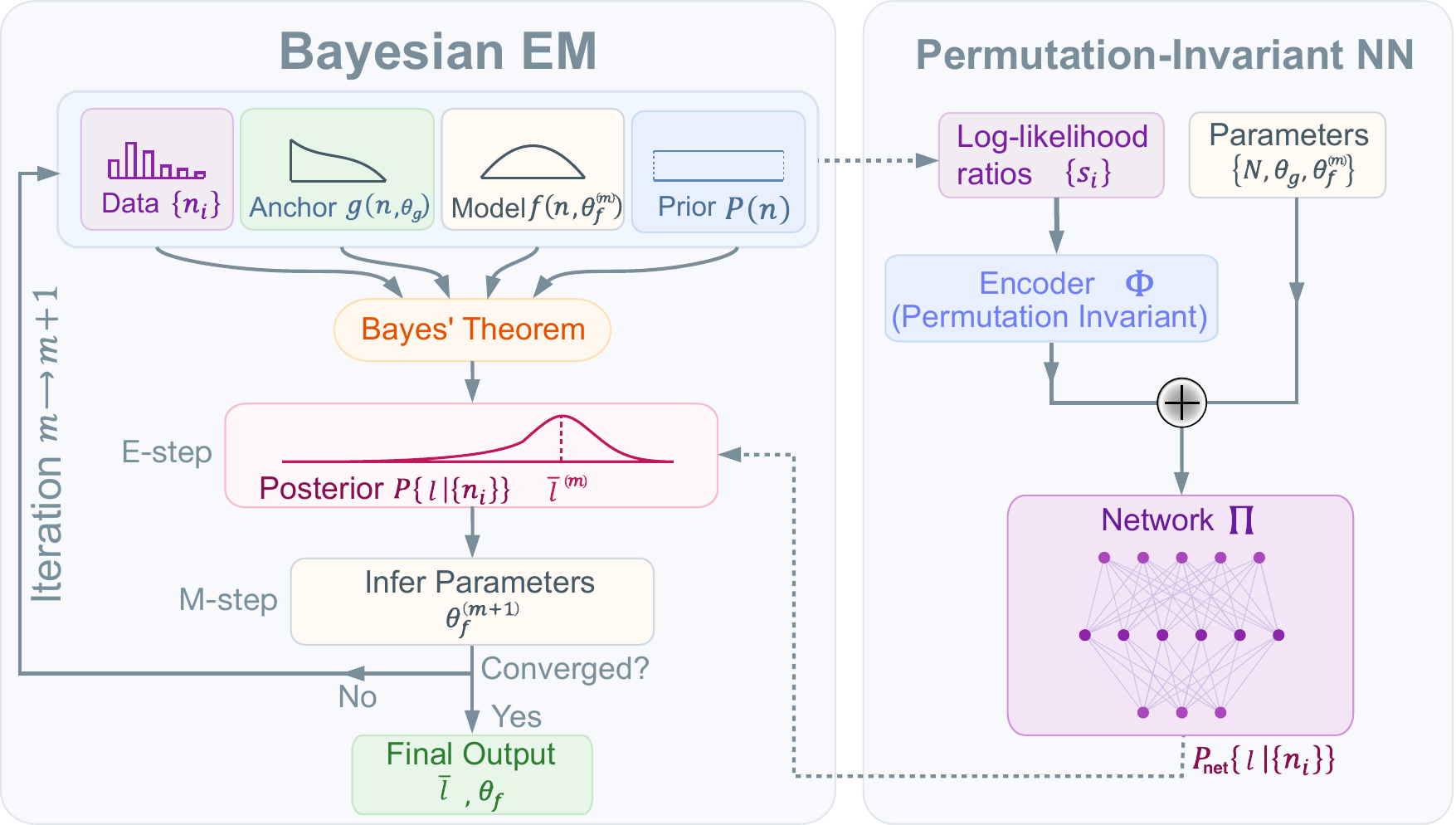}
	\caption{
		{Bayesian-EM algorithm and PI-Network architecture.} 
		Left panel: Flow chart of the weakly-anchored Bayesian-EM algorithm. The algorithm iterates between E-step (computing the posterior) and M-step (updating parameters $\theta_\text{f}$). The superscript $m$ denotes the iteration step.
		Right panel: network architecture. The observed data $\{n_i\}$ is encoded as log-likelihood ratios $\{s_i\}$ and processed through the permutation-invariant encoder $\Phi$. The encoder's output is combined with parameters $(N, \theta_\text{g}, \theta_\text{f}^m)$, then mapped through network $\Pi$ to output the posterior distribution $P_\text{net}( l |\{n_i\})$.
	}
	\label{fig:fig2}
\end{figure*}

On the other hand, when the photon count approachs to the single-photon level due to the shorten exposure time, the distribution $f(n)$ and $g(n)$ overlap with each other, resulting in the absence of bimodal structure. As an example, Fig.~\ref{fig:fig1}(b) shows the cases of $\tau = 12.5$ ms  and $\tau = 6$ ms  with single-shot photon counts $ \sim 3.1$ and $2.1$, respectively. It is evident that the threshold method ceases to be reliable for such cases. We define the histogram overlap $O(f,g) = \sum_n \sqrt{f(n)g(n)}$ to characterize the distance between $f(n)$ and $g(n)$ quantitatively, ranging from 1 (identical distributions $f=g$) to 0 (perfectly separable).

Strikingly, Bayesian inference offers a powerful means to circumvent the fundamental limitations of the threshold method. In general, both $g(n)$ and $f(n)$ are fully calibrated before the measurement, serving as the anchors. Bayes' theorem directly provides the posterior distribution $P( l |\{n_i\})$ given a sequence of observed photon counts $\{n_i | i\in[1,N]\}$:
\begin{equation}
P( l |\{n_i\}) = \frac{P( l ) \prod_{i=1}^{N} P(n_i| l )}{\int_0^1 d l  \left[P( l ) \prod_{i=1}^{N} P(n_i| l )\right]},
\label{eq:bayesian_posterior}
\end{equation}
with $P( l )$ the prior distribution. Numerically, the posterior can be computed by integration over a discrete $l$-grid to obtain the normalization (the denominator), posterior mean $\bar{ l }$, and uncertainty [the standard deviation (SD)] $\Delta l $. This is the so-called strongly anchored scenario. However, in many cases one of the states (bright or dark) is experimentally difficult to prepare or calibrate accurately beforehand.

To address such scenarios, we develop a \textit{weakly-anchored} Bayesian approach for broader applicability. In this framework, only the dark-state distribution $g(n)$ is pre-anchored, while the other $f(n;\theta_\text{f})$ remains unknown and is parametrized by $\theta_\text{f}$. Correspondly, the inference involves joint estimation of $( l , \theta_\text{f})$. We exploit the EM iteration~\cite{Moon:1996:em} which alternates between the Bayesian inference for $ l $ (E-step) and obtaining point estimates for $\theta_\text{f}$ (M-step). 

The algorithm, as schematized in Fig.~\ref{fig:fig2} (left panel), proceeds as follows: 
\begin{itemize}
\item[i)] Initializing with the prior $P( l )$, observations $\{n_i\}$, and the parameter $\theta_\text{f}^{(m)}$; 
\item[ii)] E-step: computing the posterior distribution via Eq.~(\ref{eq:bayesian_posterior}) and extracting the posterior mean $\bar{ l }^{(m)} = \int_0^1 d l   l  P( l |\{n_i\};\theta_\text{f}^{(m)})$; 
\item[iii)] M-step: obtaining the posterior weight $w_i^{(m)} = \bar{ l }^{(m)} f(n_i)/[(1-\bar{ l }^{(m)})g(n_i)+\bar{ l }^{(m)} f(n_i)]$ and performing weighted moment estimation to obtain updated $\theta_\text{f}^{(m+1)}$; 
\item[iv)] Iterating steps i)-iii) until convergence. 
\end{itemize}
The final outputs include the converged bright-state parameters $\theta_\text{f}$ and distribution $f(n; \theta_\text{f})$, as well as the coverged posterior mean $\bar{ l }$ and SD $\Delta l $. Furthermore, the relative readout fidelity can be characterized by~\cite{Myerson:2008:trapped,Elder:2020:multilevel}
\begin{equation}
F = \left[\sqrt{ l_\text{th} \bar{ l }} + \sqrt{\left(1- l_\text{th}\right)\left(1-\bar{ l }\right)}\right]^2,
\label{eq:fidelity}
\end{equation}
simply the overlap between two equal-phase states with occupations $\bar{ l }$ and $ l_\text{th}$, respectively.
Here, $ l_\text{th}$ is the reference occupation obtained through the long-exposure threshold method.

\begin{figure*}
	\includegraphics[width=0.85\textwidth]{Fig3.pdf}
	\caption{
	{Bayesian readout of Rabi oscillation.} 
	(a) Bayesian-EM inferred occupation $\bar{ l }$ versus microwave driving time $t$ for ROI 2 using the traditional Bayesian-EM algorithm, with markers correspond to exposure times $\tau \in \{25, 12.5, 6\}$ ms and error bars representing the SD $\Delta l $ of posterior distributions.
	The case of $\tau = 200$ ms (without error bar) shows the reference $ l_\text{th}$ obtained by the threshold method. The solid line and dashed lines show the fitting curves of $l_\text{th}$ and $\bar{ l }$, respectively. 
	(b) Relative readout fidelity $F$ versus exposure time $\tau$ averaged over a Rabi cycle within $t \leq 40$ $\mu$s. The horizontal line in each box represents the mean value, while the error bars indicate the data range across different $t$.
	(c) Evolution of the posterior distribution $P( l |\{n_i\})$ during iterations, starting from a uniform prior (dashed line).
	(d) Convergence trajectories of parameters $\alpha_f$ (left axis) and $\beta_f$ (right axis) versus iteration $m$.
	(e) Histograms showing dark-state count (red bars) and data count $\{n_i\}$ (blue bars), overlaid with the fitted dark-state distribution $g(n)$ and the inferred bright-state distribution $f(n;\theta_f)$.
	Data in (c)-(e) are taken at $t = 21$ $\mu$s and $\tau = 12.5$ ms with $N = 200$.
	(f) Inference time versus sample size $N$ for traditional Bayesian-EM algorithm (circles) and PI-Network-based Bayesian-EM (squares), and the acceleration ratio (diamonds, right axis), with iterations per task fixed at 10.}
	\label{fig:fig3}
\end{figure*}

It is known that Bayesian inference is intrinsically computationally expensive~\cite{Cockayne:2019:probabilistic}, demanding repeated queries to distribution functions and numerical integration in Eq.~(\ref{eq:mixture}). To accelerate the computation, we introduce a permutation-invariant neural network (PI-Network) with the associated architecture illustrated in the right panel of Fig.~\ref{fig:fig2}. It compresses the posterior computation in the E-step into a single forward propagation of the pre-trained PI-Network. Specifically, the network takes as input the log-likelihood ratios $\{s_i = \log[f(n_i)/g(n_i)]\}$ together with auxiliary parameters $(N, \theta_\text{g}, \theta_\text{f})$, where $N$ is the sample size, and $\theta_\text{g}$ and $\theta_\text{f}$ denote the fitted parameters of distributions $g$ and $f$, respectively. 
The network architecture consists of a permutation-invariant encoder $\Phi$ followed by a fully connected network $\Pi$. The encoder $\Phi$ performs order-insensitive encoding of $\{s_i\}$, meeting the requirement that the joint likelihood is invariant to the order of data. Then, the output of the encoder is aggregated with auxiliary parameters and mapped by the $\Pi$-network to the posterior distribution $P_\text{net}( l |\{n_i\})$.
The output posterior is automatically normalized due to the final softmax activation. 
Using Kullback-Leibler divergence as the loss function, the network is trained on simulated data across a wide physical parameter space \cite{SM}. 

\begin{figure*}[htbp]
	\includegraphics[width=0.8\textwidth]{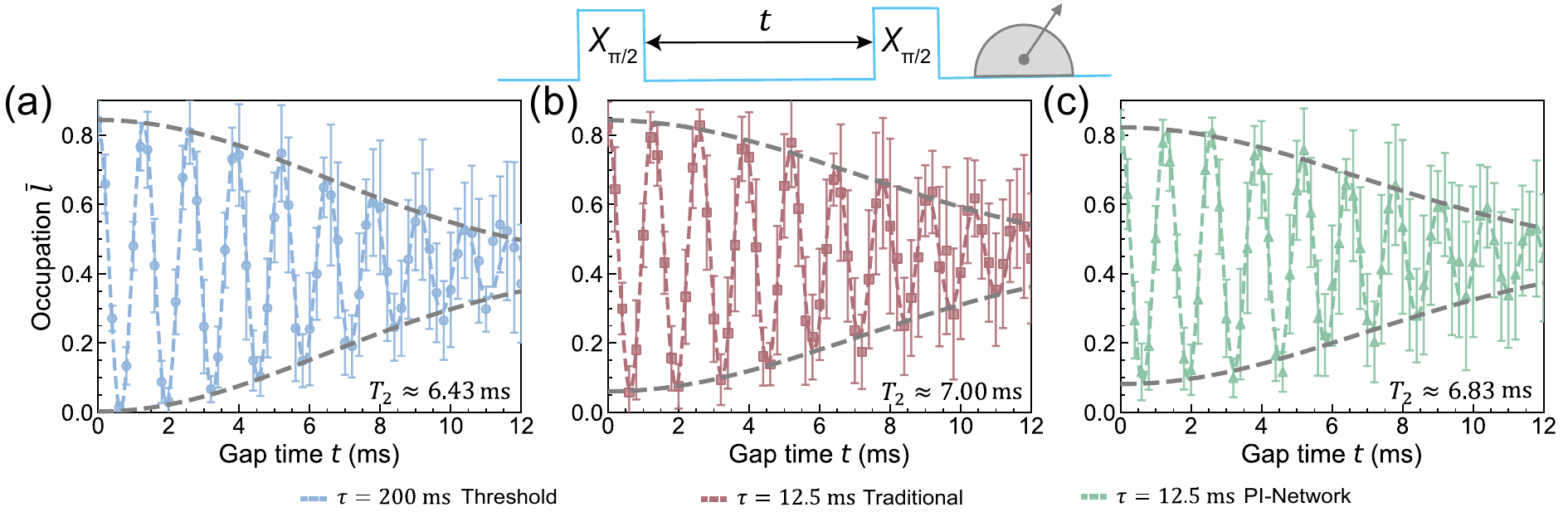}
	\caption{
	{Bayesian readout of Ramsey interferometry.} 
	Tweezer-averaged occupation $\bar{ l }$ versus free gap time $t$, with error bars denoting standard deviation across different ROIs. 
	(a) Threshold method with $\tau = 200$ ms exposure time. 
	(b) Traditional Bayesian-EM algorithm with $\tau = 12.5$ ms. 
	(c) PI-Network-based Bayesian-EM algorithm with $\tau = 12.5$ ms. 
	Dotted lines represent fits to the damped oscillation model [Eq.~(\ref{eq:ramsey})], with dashed lines indicating the Gaussian decay envelopes. 
	The extracted coherence times are $T_2 \approx 6.43$ ms, $7.00$ ms, and $6.83$ ms for panels (a), (b), and (c), respectively. 
	}
	\label{fig:fig5}
\end{figure*}

\subsection{Experimental implementation}  \label{Sec2}
We demonstrate this protocol by starting with a 4$\times$4 defect-free atom array. Here the qubit subspace is spanned by two hyberfine clock states in the ground-state manifold of $^{87}$Rb with $|0\rangle=|F=2,~m_F=0\rangle$ and $|1\rangle=|F=1,~m_F=0\rangle$. We initialize the atoms in state $|0\rangle$, and drive the transition between $|0\rangle$ and $|1\rangle$ with a resonant microwave pulse. To read out qubits via fluorescence imaging, we apply a push-out laser pulse resonant with the D2 line transition $|F=2\rangle\rightarrow|F'=3\rangle$ to remove the atoms in $|0\rangle$, and image atoms atoms remain in the $F=1$ ground-state manifold. Hence, $|0\rangle$ and $|1\rangle$ are the dark and bright state, respectively. In experiments, atomic fluorescence is collected by a moderately high numerical aperture objective (NA=0.5), and is then imaged onto an electron multiplying charge-coupled device camera (EMCCD).

We work with exposure times $\tau \in \{6,12.5,25, 200\} \text{ms}$ associated with averaged photon counts per shot $\sim \{2, 3, 18, 138\}$, respectively. The case of $\tau = 200$ ms provides threshold-method baseline. We find that both $g(n)$ and $f(n)$ exhibit over-dispersion ($\sigma^2 > \mu$), and thus can be modeled with super-Poisson distributions. The parameters $\theta = \{\alpha, \beta\}$ are related to the mean and variance by $\mu = \alpha/\beta$ and $\sigma^2 = \alpha(1+\beta)/\beta^2$ \cite{SM}. For a given exposure $\tau$, we firstly obtain $g(n)$ and its parameters $\theta_g=\{\alpha_g, \beta_g\}$ by fitting a histogram of $10^3$ background images. To read out a target state following the mixture distribution in Eq.~(\ref{eq:mixture}), we perform $N$ shots to obtain the count sequence $\{n_i\}$ that are subsequently fed into the Bayesian-EM algorithm for inference. Note that in our calculation, the prior $P( l )$ is always set to be uniform for the most general case.

We first readout the Rabi oscillation driven by the resonant microwave pulse and Fig.~\ref{fig:fig3} displays typical results for a single ROI. By varying the microwave pulse duration $t$, we prepare the qubit with different occupation $ l (t) = A\sin^2(\Omega t/2)$, where $\Omega$ and $A$ are the Rabi frequency and amplitude, respectively.
Fig.~\ref{fig:fig3}(a) displays the occupation $\bar{ l }$ inferred by the Bayesian-EM algorithm for different exposure times $\tau$, at time points $t \in \{0,7,14,21,28,35,42\}$ $\mu$s with $N = 200$, where the error bar represents the posterior SD $\Delta l $. Here all the fitted dashed lines exhibit the obvious Rabi oscillation. As a reference, $l_\text{th}$ derived from the threshold method at $\tau = 200$ ms is marked by yellow circles. The corresponding solid-line fit yields a Rabi frequency of $\Omega/2\pi \approx 25.1$ kHz and $A \approx 0.87$.
Fig.~\ref{fig:fig3}(b) gives the readout fidelity $F$ versus $\tau$ averaged over the Rabi cycle, while the error bar indicates the fluctuation in different Rabi time. The Bayesian algorithm proves to be robust and effective: $F \ge 99.99\%$ at $\tau = 200$ ms, $F = 99.78\%$ at 25 ms, $F = 99.14\%$ at 12.5 ms, and $F = 98.87\%$ at $6$ ms. 

An important measure of parameter estimation is its convergence. We thus investigate the convergence behavior of the Bayesian-EM algorithm and present the assocatied results in Fig.~\ref{fig:fig3}(c)-e. 
Taking the case of $t = 21 \mu$s and $\tau = 12.5$ ms as an example, we first examine the evolution of the posterior distribution $P( l |{n_i})$ during iteration shown in Fig.~\ref{fig:fig3}(c). The posterior peak—unlike the initial uniform prior (gray dashed line)—gradually converges to the final estimate, while its width $\Delta l $ remains largely unchanged. 
The convergence of parameters $\alpha_f$ and $\beta_f$ is shown in Fig.~\ref{fig:fig3}(d), reaching stability within about 10 iterations. Finally, the inferred distribution $f(n)$ (determined by $\theta_f$) is overlaid with the experimental histogram in Fig.~\ref{fig:fig3}(e), validating the weakly-anchored Bayesian inference despite no prior calibration of $f(n)$.
We further compute the overlap between $g(n)$ and the inferred $f(n)$, which yields $O(f,g) \approx \{0, 0.05, 0.61, 0.72\}$ for the cases of $\tau = \{200, 50, 12.5, 6\}$ ms, respectively.

As aforementioned, the Bayesian readout comes at the cost of substantial computational resources.
To accelerate the inference, we develop the PI-Network-based Bayesian inference. When benchmarked against the traditional Bayesian-EM algorithm, the PI-Network attains a drastic reduction in computation time with no loss in performance (see the SM \cite{SM} for more details). Fig.~\ref{fig:fig3}(f) compares the inference time versus sample size $N$ for both methods, with each inference performing 10 iterations \cite{note1}. 
While both methods scale linearly with $N$, their slopes differ significantly. The ratio between the two curves, defined as the speedup factor (diamonds), converges to approximately 100.56 in the large-$N$ limit, demonstrating an acceleration of nearly two orders of magnitude.

We proceed to apply our method to atomic Ramsey interferometry \cite{foot_atomic_2023}, a foundational technique in quantum information science for applications such as sensing and gate benchmarking. 
Fig.~\ref{fig:fig5} compares three measurement approaches: the threshold method with $\tau = 200$ ms (panel (a)), and the traditional and network-based Bayesian inference with $\tau = 12.5$ ms (panels (b) and (c)). 
During the free evolution, the relative phase between $|0\rangle $ and $|1\rangle $ state accumulates due to the detuned drive, while the qubit dephases. We observe that the occupation follows a damped oscillation with Gaussian decay (dashed lines):
\begin{equation}
 l (t) \sim \exp\left(-\frac{t^2}{2\sigma^2}\right) \cos(\Delta\omega t + \phi) +  l _0,
\label{eq:ramsey}
\end{equation}
where $\sigma$ characterizes the inhomogeneous coherence time $T_2$. $\Delta\omega$ is the detuning frequency between the microwave drive and atomic transition, while $\phi$ and $l_0$ are the initial phase and a offset, respectively. Using the threshold method as a reference (fitted $\Delta\omega \approx 2\pi \times 0.767$ kHz, $T_2 \approx 6.43$ ms), the traditional Bayesian-EM algorithm yields $\Delta\omega \approx 2\pi \times 0.769$ kHz and $T_2 = 7.00$ ms; and the PI-Network-based Bayesian algorithm obtains $\Delta\omega \approx 2\pi \times 0.770$ kHz and $T_2 = 6.83$ ms.

\section*{Discussion.} \label{Discussion}
We demonstrate an AI-accelerated Bayesian inference method for fluorescence readout in neutral atom arrays. This approach enables qubit readout at the single-photon level, a key benefit for suppressing photon-scattering-induced heating and atom loss. The results can be extended along several directions. 
First, photon collection efficiency-currently limited to less than $0.2\%$ by hardware such as the moderate-NA ($\sim 0.5$), lower-transmission ($\sim 28\%$ at 780 nm) objective  and other optical losses-can be substantially improved. Upgrading to higher-performance optics (e.g., objective from Special Optics with $\sim 94\%$ transmission and NA $\sim 0.65$) would boost efficiency to $\sim 5.2\%$, enabling single-photon-level readout with exposure times as short as  $\sim 200$ $\mu$s.
Secondly, integrating Bayesian inference directly into measurement hardware presents a promising direction for further optimization. This would enable real-time adaptive schemes~\cite{Lee:2022:adaptive,Fiderer:2021:neuralnetwork}, where the posterior from one measurement cycle informs the prior for the next, thereby enhancing statistical efficiency. Key challenges include compressing inference time to the millisecond scale, requiring more efficient network architectures. Conversely, the inherent parallelizability of neural networks allows for the simultaneous readout of large-scale atom arrays via a single forward pass of pretrained networks~\cite{MartinezDorantes:2017:parallel}.

The weakly-anchored Bayesian scheme offers practical advantages in situations where calibrating the two state distributions differs in difficulty—a scenario ubiquitous across quantum platforms.
In our setup, the dark-state distribution $g(n)$ is easily anchored as it coincides with the background photon-count distribution of an empty tweezer; whereas anchoring the bright-state $f(n)$ would require near-unit preparation fidelity of $|1\rangle$, which is more challenging due to the need for precise microwave pulse calibration.
Similar situations appear in solid-state spin systems such as quantum dots and nitrogen-vacancy centers \cite{vukusicSingleShotReadoutHole2018,braithwaiteMultipleSuperconductingPhases2019}, where the readout window is often comparable to the relaxation time $T_1$. Consequently, spins initialized in the excited state may relax during readout, rendering the effective "bright" histogram time-dependent and nonstationary, whereas the ground-state distribution remains stable and easier to calibrate.

\begin{acknowledgments}
This work is supported by NNSFC under Grant Nos. U25A20198 and 12574296, and the Quantum Science and Technology-National Science and Technology Major Project  under Grant No. 2023ZD0300900.
\end{acknowledgments}

\bibliography{Ref}

@misc{SM,
  note={See the Supplemental Materials for more details}
}

@misc{note1,
  note={All computations are performed on a MacBook Studio equipped with an Apple M2 Max chip and 96 GB shared memory.}
}

@article{braithwaiteMultipleSuperconductingPhases2019,
  title={Multiple Superconducting Phases in a Nearly Ferromagnetic System},
  author={Braithwaite, D. and Vališka, M. and Knebel, G. and Lapertot, G. and Brison, J.-P. and Pourret, A. and Zhitomirsky, M. E. and Flouquet, J. and Honda, F. and Aoki, D.},
  journal={Communications Physics},
  volume={2},
  number={1},
  pages={147},
  year={2019},
  doi={10.1038/s42005-019-0248-z},
  publisher={Nature Publishing Group}
}

@article{Heisenberg1927,
  title={Über den anschaulichen Inhalt der quantentheoretischen Kinematik und Mechanik},
  author={Heisenberg, W},
  journal={Z. Physik},
  volume={43},
  pages={172--198},
  year={1927},
  doi={10.1007/BF01397280},
}

@article{GerlachStern1922,
  title={Der experimentelle Nachweis der Richtungsquantelung im Magnetfeld},
  author={Gerlach, W. and Stern, O.},
  journal={Z. Physik},
  volume={9},
  pages={349--352},
  year={1922},
  doi={10.1007/BF01397280},
}

@article{Castelvecchi2022,
  title={The Stern–Gerlach experiment at 100},
  author={Castelvecchi, D},
  journal={Nature Review Physics},
  volume={4},
  number={3},
  pages={140--142},
  year={2022},
  doi={10.1038/s42254-022-00436-4},
  publisher={Nature Publishing Group}
}

@article{Bell1964,
  title={On the Einstein-Podolsky-Rosen paradox},
  author={Bell, J. S},
  journal={Physics},
  volume={1},
  number={3},
  pages={195--200},
  year={1964},
  doi={10.1103/PhysicsPhysiqueFizika.1.195},
}

@article{vukusicSingleShotReadoutHole2018,
  title={Single-{{Shot Readout}} of {{Hole Spins}} in {{Ge}}},
  author={Vukušić, Lada and Kukučka, Josip and Watzinger, Hannes and Milem, Joshua Michael and Schäffler, Friedrich and Katsaros, Georgios},
  journal={Nano Letters},
  volume={18},
  number={11},
  pages={7141--7145},
  year={2018},
  doi={10.1021/acs.nanolett.8b03217},
  publisher={American Chemical Society}
}

@article{Zeilinger:1999:foundations,
  title={Experiment and the foundations of quantum physics},
  author={Zeilinger, Anton},
  journal={Reviews of Modern Physics},
  volume={71},
  number={2},
  pages={S288--S297},
  year={1999},
  doi={10.1103/RevModPhys.71.S288},
  publisher={APS}
}

@article{Aspect:1982:bell,
  title={Experimental test of {Bell's} inequalities using time-varying analyzers},
  author={Aspect, Alain and Dalibard, Jean and Roger, Gérard},
  journal={Physical Review Letters},
  volume={49},
  number={25},
  pages={1804--1807},
  year={1982},
  doi={10.1103/PhysRevLett.49.1804},
  publisher={APS}
}

@article{Clauser:1972:bell,
  title={Experimental Test of Local Hidden-Variable Theories},
  author={Freedman, S. J. and Clauser, J. F.},
  journal={Physical Review Letters},
  volume={28},
  number={14},
  pages={938--941},
  year={1972},
  doi={10.1103/PhysRevLett.28.938},
  publisher={APS}
}

@article{Zelinger:1998:bell,
  title={Experimental Entanglement Swapping: Entangling Photons That Never Interacted},
  author={Pan, J.-W. and Bouwmeester, D. and Weinfurter, H. and Zeilinger, A.},
  journal={Physical Review Letters},
  volume={80},
  number={18},
  pages={3891--3894},
  year={1998},
  doi={10.1103/PhysRevLett.80.3891},
  publisher={APS}
}

@article{Briegel:2009:measurement,
  title={Measurement-based quantum computation},
  author={Briegel, Hans J. and Browne, Daniel E. and Dür, Wolfgang and Raussendorf, Robert and Van den Nest, Maarten},
  journal={Nature Physics},
  volume={5},
  number={1},
  pages={19--26},
  year={2009},
  doi={10.1038/nphys1157},
  publisher={Nature Publishing Group}
}

@article{TerhalQEC2015,
  title={Quantum error correction for quantum memories},
  author={Terhal, Barbara M. },
  journal={Review of Modern Physics},
  volume={87},
  number={2},
  pages={307--346},
  year={2015},
  doi={10.1103/RevModPhys.87.307},
  publisher={APS}
}

@article{CaiRMPQEM2023,
  title = {Quantum error mitigation},
  author = {Cai, Zhenyu and Babbush, Ryan and Benjamin, Simon C. and Endo, Suguru and Huggins, William J. and Li, Ying and McClean, Jarrod R. and O'Brien, Thomas E.},
  journal = {Rev. Mod. Phys.},
  volume = {95},
  issue = {4},
  pages = {045005},
  numpages = {37},
  year = {2023},
  publisher = {American Physical Society},
  doi = {10.1103/RevModPhys.95.045005},
}

@article{Broweays2020NP,
  title={Many-body physics with individually controlled Rydberg atoms},
  author={Antoine Browaeys and Thierry Lahaye},
  journal={Nature Physics},
  volume={16},
  number={2},
  pages={132--142},
  year={2020},
  doi={10.1038/s41567-019-0733-z},
  publisher={Nature Publishing Group}
}

@article{KlausRMP2010,
  title = {Quantum information with Rydberg atoms},
  author = {Saffman, M. and Walker, T. G. and Mølmer, K.},
  journal = {Review of Modern Physics},
  volume = {82},
  issue = {3},
  pages = {2313--2363},
  numpages = {0},
  year = {2010},
  publisher = {American Physical Society},
  doi = {10.1103/RevModPhys.82.2313},
}

@article{Urban:2009:rydberg,
  title={Observation of {Rydberg} blockade between two atoms},
  author={Urban, E. and Johnson, T. A. and Henage, T. and Isenhower, L. and Yavuz, D. D. and Walker, T. G. and Saffman, M.},
  journal={Nature Physics},
  volume={5},
  number={2},
  pages={110--114},
  year={2009},
  doi={10.1038/nphys1178},
  publisher={Nature Publishing Group}
}

@article{Gaetan:2009:rydberg,
  title={Observation of collective excitation of two individual atoms in the {Rydberg} blockade regime},
  author={Gaëtan, A. and Miroshnychenko, Y. and Wilk, T. and Chotia, A. and Viteau, M. and Comparat, D. and Pillet, P. and Browaeys, A. and Grangier, P.},
  journal={Nature Physics},
  volume={5},
  number={2},
  pages={115--118},
  year={2009},
  doi={10.1038/nphys1183},
  publisher={Nature Publishing Group}
}

@article{Moon:1996:em,
  title={The expectation-maximization algorithm},
  author={Moon, Todd K.},
  journal={IEEE Signal Processing Magazine},
  volume={13},
  number={6},
  pages={47--60},
  year={1996},
  doi={10.1109/79.543975},
  publisher={IEEE}
}

@article{Myerson:2008:trapped,
  title={High-fidelity readout of trapped-ion qubits},
  author={Myerson, A. H. and Szwer, D. J. and Webster, S. C. and Allcock, D. T. C. and Curtis, M. J. and Imreh, G. and Sherman, J. A. and Stacey, D. N. and Steane, A. M. and Lucas, D. M.},
  journal={Physical Review Letters},
  volume={100},
  number={20},
  pages={200502},
  year={2008},
  doi={10.1103/PhysRevLett.100.200502},
  publisher={APS}
}

@article{manetsch2025tweezer,
  title={A tweezer array with 6,100 highly coherent atomic qubits},
  author={Manetsch, Hannah J and Nomura, Gyohei and Bataille, Elie and Lv, Xudong and Leung, K H and Endres, Manuel},
  journal={Nature},
  volume={647},
  pages={60},
  year={2025},
  publisher={Nature Publishing Group}
}

@article{Sherson:2010:fluorescence,
  title={Single-atom-resolved fluorescence imaging of an atomic {Mott} insulator},
  author={Sherson, Jacob F. and Weitenberg, Christof and Endres, Manuel and Cheneau, Marc and Bloch, Immanuel and Kuhr, Stefan},
  journal={Nature},
  volume={467},
  number={7311},
  pages={68--72},
  year={2010},
  doi={10.1038/nature09378},
  publisher={Nature Publishing Group}
}

@article{WongCampos:2016:imaging,
  title={High-resolution adaptive imaging of a single atom},
  author={Wong-Campos, J. D. and Johnson, K. G. and Neyenhuis, B. and Mizrahi, J. and Monroe, C.},
  journal={Nature Photonics},
  volume={10},
  number={10},
  pages={606--610},
  year={2016},
  doi={10.1038/nphoton.2016.136},
  publisher={Nature Publishing Group}
}

@article{Lis:2023:midcircuit,
  title={Midcircuit operations using the omg architecture in neutral atom arrays},
  author={Lis, Joanna W. and Senoo, Aruku and McGrew, William F. and Rönchen, Felix and Jenkins, Alec and Kaufman, Adam M.},
  journal={Physical Review X},
  volume={13},
  number={4},
  pages={041035},
  year={2023},
  doi={10.1103/PhysRevX.13.041035},
  publisher={APS}
}

@article{Graham:2023:midcircuit,
  title={Midcircuit measurements on a single-species neutral alkali atom quantum processor},
  author={Graham, T. M. and Phuttitarn, L. and Chinnarasu, R. and Song, Y. and Poole, C. and Jooya, K. and Scott, J. and Scott, A. and Eichler, P. and Saffman, M.},
  journal={Physical Review X},
  volume={13},
  number={4},
  pages={041051},
  year={2023},
  doi={10.1103/PhysRevX.13.041051},
  publisher={APS}
}

@article{Endres:2016:assembly,
  title={Atom-by-atom assembly of defect-free one-dimensional cold atom arrays},
  author={Endres, Manuel and Bernien, Hannes and Keesling, Alexander and Levine, Harry and Anschuetz, Eric R. and Krajenbrink, Alexandre and Senko, Crystal and Vuletic, Vladan and Greiner, Markus and Lukin, Mikhail D.},
  journal={Science},
  volume={354},
  number={6315},
  pages={1024--1027},
  year={2016},
  doi={10.1126/science.aah3752},
  publisher={AAAS}
}

@article{Barredo:2016:assembler,
  title={An atom-by-atom assembler of defect-free arbitrary two-dimensional atomic arrays},
  author={Barredo, Daniel and de Léséleuc, Sylvain and Lienhard, Vincent and Lahaye, Thierry and Browaeys, Antoine},
  journal={Science},
  volume={354},
  number={6315},
  pages={1021--1023},
  year={2016},
  doi={10.1126/science.aah3778},
  publisher={AAAS}
}

@article{Ahn:2016:assembler,
  title={In situ single-atom array synthesis using dynamic holographic optical tweezers},
  author={Kim, H. and Lee, W. and Lee, H.-G. and Jo, H. and Song, Y.and Ahn, J.},
  journal={Nature Communications},
  volume={7},
  pages={13317},
  year={2016},
  doi={10.1038/ncomms13317},
  publisher={ature Publishing Group}
}

@misc{wurtz2023aquilaqueras256qubitneutralatom,
      title={Aquila: QuEra's 256-qubit neutral-atom quantum computer}, 
      author={Jonathan Wurtz and Alexei Bylinskii and Boris Braverman and Jesse Amato-Grill and Sergio H. Cantu and Florian Huber and Alexander Lukin and Fangli Liu and Phillip Weinberg and John Long and Sheng-Tao Wang and Nathan Gemelke and Alexander Keesling},
      year={2023},
      eprint={2306.11727},
      archivePrefix={arXiv},
      primaryClass={quant-ph},
      url={https://arxiv.org/abs/2306.11727}, 
}

@article{Su:2025:imaging,
  title={Fast single atom imaging for optical lattice arrays},
  author={Su, Lin and Douglas, Alexander and Szurek, Michal and Hébert, Alexandre and Krahn, Aaron and Cao, Yunfan and Glidden, Jeremy A. H. and Heil, Thomas and Mackoit-Sinkevičienė, Milda and Çuhadar, Emre and Bilitewski, Thomas and Kubica, Aleksander and Rey, Ana Maria and Bakr, Waseem S. and Greiner, Markus},
  journal={Nature Communications},
  volume={16},
  number={1},
  pages={814},
  year={2025},
  doi={10.1038/s41467-025-56305-y},
  publisher={Nature Publishing Group}
}

@article{Phuttitarn:2024:machinelearning,
  title={Enhanced measurement of neutral-atom qubits with machine learning},
  author={Phuttitarn, L. and Becker, B. M. and Chinnarasu, R. and Graham, T. M. and Kwon, M. and Saffman, M.},
  journal={Physical Review Applied},
  volume={22},
  number={2},
  pages={024011},
  year={2024},
  doi={10.1103/PhysRevApplied.22.024011},
  publisher={APS}
}

@article{VonToussaint:2011:bayesian,
  title={{Bayesian} inference in physics},
  author={Von Toussaint, Udo},
  journal={Reviews of Modern Physics},
  volume={83},
  number={3},
  pages={943--999},
  year={2011},
  doi={10.1103/RevModPhys.83.943},
  publisher={APS}
}

@article{VanDeSchoot:2021:bayesian,
  title={{Bayesian} statistics and modelling},
  author={Van de Schoot, Rens and Depaoli, Sarah and King, Ruth and Kramer, Bianca and Märtens, Kaspar and Tadesse, Mahlet G. and Vannucci, Marina and Gelman, Andrew and Veen, Duco and Willemsen, Joukje and Yau, Christopher},
  journal={Nature Reviews Methods Primers},
  volume={1},
  number={1},
  pages={1},
  year={2021},
  doi={10.1038/s43586-020-00001-2},
  publisher={Nature Publishing Group}
}

@article{Fuchs:2013:qbism,
  title={Quantum-{Bayesian} coherence},
  author={Fuchs, Christopher A. and Schack, Rüdiger},
  journal={Reviews of Modern Physics},
  volume={85},
  number={4},
  pages={1693--1715},
  year={2013},
  doi={10.1103/RevModPhys.85.1693},
  publisher={APS}
}

@inproceedings{Zaheer:2017:deepsets,
  title={Deep sets},
  author={Zaheer, Manzil and Kottur, Satwik and Ravanbakhsh, Siamak and Poczos, Barnabas and Salakhutdinov, Ruslan and Smola, Alexander J.},
  booktitle={Advances in Neural Information Processing Systems},
  volume={30},
  pages={3391--3401},
  year={2017},
  publisher={Curran Associates, Inc.}
}

@inproceedings{Lee:2019:settransformer,
  title={Set transformer: A framework for attention-based permutation-invariant neural networks},
  author={Lee, Juho and Lee, Yoonho and Kim, Jungtaek and Kosiorek, Adam and Choi, Seungjin and Teh, Yee Whye},
  booktitle={International Conference on Machine Learning},
  pages={3744--3753},
  year={2019},
  publisher={PMLR}
}

@article{Elder:2020:multilevel,
  title={High-fidelity measurement of qubits encoded in multilevel superconducting circuits},
  author={Elder, S. S. and Wang, C. S. and Reinhold, P. and Hann, C. T. and Chou, K. S. and Lester, B. J. and Rosenblum, S. and Frunzio, L. and Jiang, L. and Schoelkopf, R. J.},
  journal={Physical Review X},
  volume={10},
  number={1},
  pages={011001},
  year={2020},
  doi={10.1103/PhysRevX.10.011001},
  publisher={APS}
}

@article{Todaro:2021:superconducting,
  title={State readout of a trapped ion qubit using a trap-integrated superconducting photon detector},
  author={Todaro, S. L. and Verma, V. B. and McCormick, K. C. and Allcock, D. T. C. and Mirin, R. P. and Wineland, D. J. and Nam, S. W. and Wilson, A. C. and Leibfried, D. and Slichter, D. H.},
  journal={Physical Review Letters},
  volume={126},
  number={1},
  pages={010501},
  year={2021},
  doi={10.1103/PhysRevLett.126.010501},
  publisher={APS}
}

@article{Lee:2022:adaptive,
  title={Quantum-inspired multi-parameter adaptive {Bayesian} estimation for sensing and imaging},
  author={Lee, K. K. and Gagatsos, C. N. and Guha, S. and Ashok, A.},
  journal={IEEE Journal of Selected Topics in Signal Processing},
  volume={17},
  number={2},
  pages={491--505},
  year={2022},
  doi={10.1109/JSTSP.2022.3214774},
  publisher={IEEE}
}

@article{Fiderer:2021:neuralnetwork,
  title={Neural-network heuristics for adaptive {Bayesian} quantum estimation},
  author={Fiderer, Lukas J. and Schuff, Jonas and Braun, Daniel},
  journal={PRX Quantum},
  volume={2},
  number={2},
  pages={020303},
  year={2021},
  doi={10.1103/PRXQuantum.2.020303},
  publisher={APS}
}

@article{MartinezDorantes:2017:parallel,
  title={Fast nondestructive parallel readout of neutral atom registers in optical potentials},
  author={Martinez-Dorantes, M. and Alt, W. and Gallego, J. and Ghosh, S. and Ratschbacher, L. and Völzke, Y. and Meschede, D.},
  journal={Physical Review Letters},
  volume={119},
  number={18},
  pages={180503},
  year={2017},
  doi={10.1103/PhysRevLett.119.180503},
  publisher={APS}
}

@article{Cockayne:2019:probabilistic,
  title={{Bayesian} probabilistic numerical methods},
  author={Cockayne, Jon and Oates, Chris J. and Sullivan, Tim J. and Girolami, Mark},
  journal={SIAM Review},
  volume={61},
  number={4},
  pages={756--789},
  year={2019},
  doi={10.1137/17M1139357},
  publisher={SIAM}
}

@article{bluvstein2024logical,
  title={Logical quantum processor based on reconfigurable atom arrays},
  author={Bluvstein, Dolev and Evered, Simon J and Geim, Alexandra A and Li, Jian S and Zhou, Hengyun and Manovitz, Tom and Ebadi, Sepehr and others},
  journal={Nature},
  volume={626},
  number={7997},
  pages={58--65},
  year={2024},
  doi={10.1038/s41586-023-06927-3},
  publisher={Nature Publishing Group},
}

@article{bluvstein2025fault,
  title={A fault-tolerant neutral-atom architecture for universal quantum computation},
  author={Bluvstein, Dolev and Geim, Alexandra A and Li, Sophie H and Evered, Simon J and others},
  journal={Nature},
  year={2025},
  doi={10.1038/s41586-025-09848-5},
  publisher={Nature Publishing Group},
}

@article{zhou2025low,
  title={Low-overhead transversal fault tolerance for universal quantum computation},
  author={Zhou, Hengyun and Zhao, C and Cain, Madelyn and others},
  journal={Nature},
  year={2025},
  publisher={Nature Publishing Group},
  doi={10.1038/s41586-025-09543-5},

}

@article{Qiao2025RydbergQS,
  title={Realization of a doped quantum antiferromagnet in a Rydberg tweezer array},
  author={Qiao, M. and Emperauger, G. and Chen, C. and others},
  journal={Nature},
  volume={644},
  number={8078},
  pages={889--895},
  year={2025},
  doi={10.1038/s41586-025-09377-1},
  publisher={Nature Publishing Group},
}

@article{Chen2025RydbergQS,
author = {Cheng Chen  and Gabriel Emperauger  and Guillaume Bornet  and Filippo Caleca  and Bastien Gély  and Marcus Bintz  and Shubhayu Chatterjee  and Vincent Liu  and Daniel Barredo  and Norman Y. Yao  and Thierry Lahaye  and Fabio Mezzacapo  and Tommaso Roscilde  and Antoine Browaeys },
title = {Spectroscopy of elementary excitations from quench dynamics in a dipolar XY Rydberg simulator},
journal = {Science},
volume = {389},
number = {6759},
pages = {483--487},
year = {2025},
doi = {10.1126/science.adn0618},
}

@article{ma2023mid-circuit,
  title={High-fidelity gates and mid-circuit erasure conversion in an atomic qubit},
  author={Ma, Shuo and Liu, Genyue and Peng, Pai and Zhang, Bichen and Jandura, Sven and Claes, Jahan and  Burgers, Alex P. and Pupillo, Guido and Puri, Shruti and Thompson, Jeff D.},
  journal={Nature},
  volume={622},
  number={7982},
  pages={279--284},
  year={2023},
  doi={10.1038/s41586-023-06438-1},
  publisher={Nature Publishing Group},

}

@book{foot_atomic_2023,
	address = {Oxford},
	title = {Atomic physics},
	isbn = {9780198506959 9781383021462},
	publisher = {Oxford University Press},
	author = {Foot, C. J.},
	year = {2023},
}

\end{document}